# Exploiting the Generative Adversarial Network Approach to Create a Synthetic Topography Corneal Image

**Samer Kais Jameel** [1], **Sezgin Aydin** [2], **Nebras H. Ghaeb** [3], **Jafar Majidpour** [1], **Tarik A. Rashid** [4,*], **Sinan Q. Salih** [5] and **P. S. JosephNg** [6,*]

[1] Computer Science Department, University of Raparin, Rania 46012, Iraq. samer.kais@uor.edu.krd
[2] Department of Natural and Mathematical Sciences, Engineer Faculty, Tarsus University, Tarsus 33402, Turkey. sezginaydin@tarsus.edu.tr
[3] Biomedical Engineering Department, Al-Khawarezmi Eng. College, University of Baghdad, Baghdad 1001, Iraq. nebras@kecbu.uobaghdad.edu.iq; jafar.majidpoor@uor.edu.krd
[4] Computer Science and Engineering Department, University of Kurdistan Hewlêr, Erbil 44001, Iraq. tarik.ahmed@ukh.edu.krd
[5] Department of Communication Technology Engineering, College of Information Technology, Imam Ja'afar Al-Sadiq University, Baghdad 10011, Iraq. sinan.salih@sadiq.edu.iq
[6] Faculty of Data Science & Information Technology, INTI International University, Persiaran Perdana BBN, Nilai 71800, Negeri Sembilan, Malaysia. joseph.ng@newinti.edu.my
* Correspondence: tarik.ahmed@ukh.edu.krd (T.A.R.); joseph.ng@newinti.edu.my (p.S.J.)

**Abstract:** Corneal diseases are the most common eye disorders. Deep learning techniques are used to perform automated diagnoses of cornea. Deep learning networks require large-scale annotated datasets, which is conceded as a weakness of deep learning. In this work, a method for synthesizing medical images using conditional generative adversarial networks (CGANs), is presented. It also illustrates how produced medical images may be utilized to enrich medical data, improve clinical decisions, and boost the performance of the conventional neural network (CNN) for medical image diagnosis. The study includes using corneal topography captured using a Pentacam device from patients with corneal diseases. The dataset contained 3448 different corneal images. Furthermore, it shows how an unbalanced dataset affects the performance of classifiers, where the data are balanced using the resampling approach. Finally, the results obtained from CNN networks trained on the balanced dataset are compared to those obtained from CNN networks trained on the imbalanced dataset. For performance, the system estimated the diagnosis accuracy, precision, and F1-score metrics. Lastly, some generated images were shown to an expert for evaluation and to see how well experts could identify the type of image and its condition. The expert recognized the image as useful for medical diagnosis and for determining the severity class according to the shape and values, by generating images based on real cases that could be used as new different stages of illness between healthy and unhealthy patients.

## 1. Introduction

Medical image datasets are one of the most important problems facing researchers in the field of machine learning [1]. The limited amount of medical data comes from the difficulty of capturing it [2]. With the problem of final ethical approval, the acquisition and labelling of medical images are time-consuming, and considerable effort needs to be spent by both researchers and specialists [3,4]. Several studies tried to overcome the dataset scarcity challenge through the famous task in computer vision, a method called data augmentation [5]. Using classic data augmentation can give a simple extra feature where it involves simple modifications, such as rotation, translation, scaling, and flipping [6]. On the other hand, some researchers employed innovative techniques for data augmentation to improve the system training process, based on synthesizing high-quality sample images using a generative model known as generative adversarial networks (GANs) [7–9].

The GANs involved two networks; the first generates a real image from the input with the help of the noise, and the other discriminates between real and fake (generated by the first network) images. This model has been used in many studies hoping to generate realistic images, especially for medical imaging applications, such as image-to-image translation [10], image inpainting [11], segmentation-to-image translation [12], medical cross-modality translations [13], and label-to-segmentation translation [14].

Exploiting the GAN models by researchers led to the creation of cross-modality images, such as a PET scan, which was generated from a CT scan of the abdomen to show the presence of liver lesions. The GAN model of image inpainting has served as inspiration for many studies. Costa et al. [15] used a fully convolutional network to learn retinal vessel segmentation images. The binary vessel tree was then translated into a new retinal image. By using chest X-ray images, Dai et al. [16] generated lung and heart image segmentation by training a GAN model. Xu et al. [17] trained a model to translate brain MRI images into binary segmentation maps for brain tumour images. Nie et al. [18] trained a patch-based GAN to translate between brain CT and MRI images. As a step of image refinement, they recommended using an auto-context model. Schlegl et al. [19] trained a GAN model on normal retinal. To detect anomalies in retinal images, the model was tested on normal and abnormal data.

Based on what was mentioned above, the scarcity of data needs to be resolved so that researchers can use it more freely to analyze that data and produce results that serve the scientific process. The latter motivated the authors of this paper to use GAN models with the ability to synthesize real images, increase the existing data, and overcome the problem of lacking data. In this work, high-quality corneal images based on GAN models are synthesized for a specific task of corneal disease diagnosis to improve the clinical decision by introducing different stages and predicted shapes for images with illness. As an illustrated sample of manipulation for the imaging in the cornea, the different stages of keratoconus are, in most cases, unclear in borderlines. From a clinical perspective, overlapping features between stages of keratoconus lead to a controversial approach to treatment. To decide the severity and clinical or surgical procedure of work per patient clinically, considerable evidence is collected from different images per case to reach the final approach. The possibility of studying the effect and weight of this evidence per case is an attractive medical training to produce a final highly medical sensation and observation for the trained physician. In more detail, thinning in pachymetry images with its location, steepening in the inferior or superior position of the tangential mapping, and the isolated land or tongue shape that may appear in elevation front and back maps, with the astigmatism axis and obliqueness of the bowtie, would improve the effectiveness of the final diagnosis.

The cornea, which protects the eye from external substances and helps to control visual focus, is stiff but very sensitive to touch [20]. There are many corneal disorders, for instance, bullous keratopathy, Cogan syndrome, corneal ulcer, herpes simplex keratitis, herpes zoster ophthalmicus, etc. [21]. Any disorders in the cornea may cause ripping, discomfort, and dwindling vision clarity and, finally, may lead to blindness. On the other hand, any action on the cornea, such as vision correction, requires a diagnosis of the cornea's health before treatment [22]. Clinical decisions on the human cornea require reviewing numerous aspects, and ophthalmologists must handle this revision. Corneal topographical parameters are so extensive that it is difficult for surgeons or ophthalmologists to remember them all and make decisions [23]. As a consequence, based on deep learning models, we also proposed to build a sophisticated medical system using the original and the generated images (using the GAN model) for diagnosing corneal cases, to aid clinicians in the interpretation of medical images and improve clinical decision-making.

Many researchers used a variety of complex and diverse medical devices to collect data, as well as a variety of diagnostic approaches. Salih and Hussein (2018) used 732 submaps as inputs to the deep learning network; a kind of deep learning technology called the VGG-16 network was utilized to predict corneal abnormalities and normality [24]. The detection of the keratoconus eyes dataset and recognition of the normal cornea was the focus of a group of authors who used 145 normal cornea cases and 312 keratoconus cases from a database of photographs. As a classification tool, they used support vector machine (SVM) and multilayer perceptron methods. The features were extracted from the input images, then passed to the classifiers [25]. A group of researchers used a compilation of data from both Placido and Scheimpug as a feature vector. The prototype was tested with and without a posterior corneal surface, and it performed well in both situations. The thickness and posterior characteristics were found to be critical tools for predicting corneal keratoconus and avoiding corneal ectasia surgery in patients with early corneal ectasia disease [26]. Researchers employed machine learning techniques, such as ANN, SVM, regression analysis, and decision tree algorithms, to identify the disease. The information was gathered from a group of patients; in total, 23 cases of ectasia after LASIK were discovered, as well as 266 stable post-LASIK cases with over a year of follow-up. They concluded that this study method still needed to be validated [27]. Samer et al. presented a method known as SWFT for diagnosing the corneal image by extracting features from the corneal image using a Wavelet and diagnosing it using an SVM classifier [28]. In 2021, Samer and his participants designed an LIP algorithm to extract corneal image features, and they evaluated their method using many classifiers. Thus, they could train a system capable of automatically classifying corneal diseases [22]. We used deep learning techniques in the current study to diagnose corneal diseases. GAN networks were used as a tool to generate realistic corneal images. On the other hand, pre-trained convolutional neural networks (CNN) [29–31] are employed in diagnosing corneal diseases, which have recently been used in many medical imaging studies and have been reported to improve performance for a broad range of medical tasks.

This paper has made the following contributions:

(1) Using the GAN model for creating high-quality corneal images from topographical images to solve the scarcity of the cornea dataset.

(2) Examining various transfer learning methods as a based solution for the corneal diagnosis task.

(3) Augmentation of the dataset to be used in training the networks, using the generated synthetic data for improved clinical decisions.

(4) Solving the issue of time consumption that is suffered by deep learning networks.

## 2. Corneal Diseases Diagnosis

This section begins by describing the data and its features. The architecture of the GAN model for cornea image creation is discussed after that. Due to the restricted quantity of data available for training transfer learning networks, we have presented a method for augmenting synthesized images.

*2.1. Dataset*

The dataset is made up of images taken by scanning the cornea with a device called Pentacam, which generates various images and parameters known as corneal topography. Ophthalmologists use corneal topography to check eye conditions in clinics. Each patient's eye data, which includes four corneal maps (sagittal, corneal thickness (CT), elevation front (EF), and elevation back maps (EB)) with a set of parameters, are saved independently [32] (see Figure 1). The data were gathered using a Pentacam (OCULUS, Germany), an image Scheimpflug instrument. The camera scans the eye from many angles in a circular pattern, producing maps with information about the anterior and posterior parts of the cornea and a quick screening report. The Pentacam can be upgraded and altered to meet the user's requirements [33].

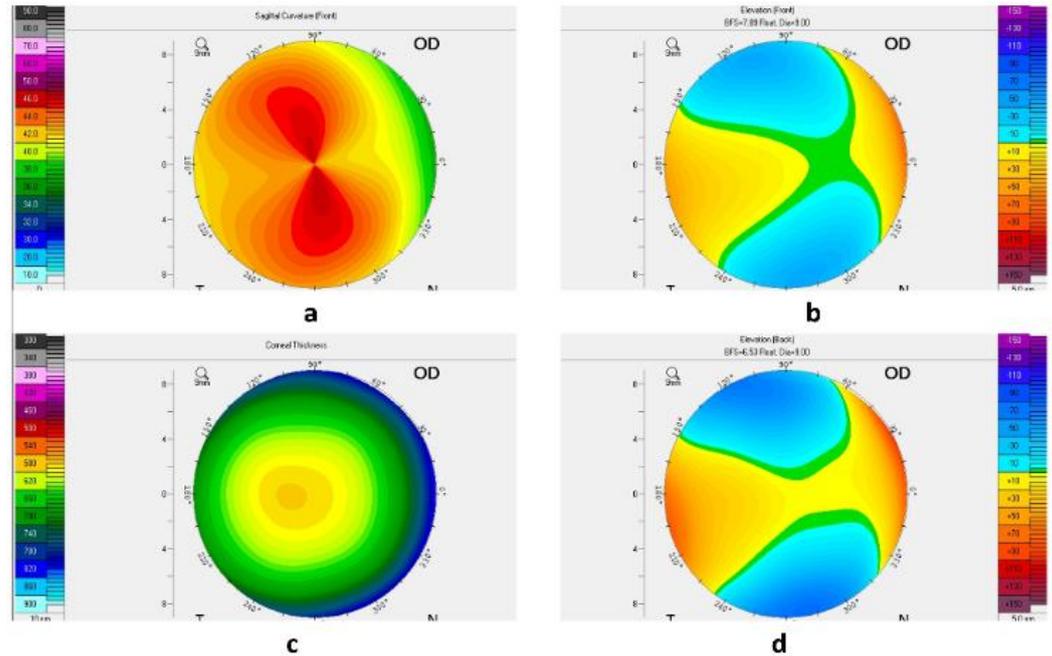

**Figure 1.** The four corneal maps: (**a**) sagittal, (**b**) elevation front, (**c**) corneal thickness, and (**d**) elevation back maps.

It is worth noting that the data were obtained from the Al-Amal center in Baghdad, Iraq, and the data were labelled with the help of eye specialists, Dr. Nebras H. Gareb, an Ophthalmic Consultant, and Dr. Sohaib A. Mohammed and Dr. Ali A. Al-Razaq, Senior Specialist Ophthalmologists. The images were categorized based on all four corneal maps, and each map was treated separately and labelled as normal or abnormal. As such, we have eight categories of cornea cases. The collected data contains 3448 images of the four maps that have been scientifically collected and classified. The number of images for each class is 248 Normal_Sagittal, 460 Abnormal_Sagittal, 338 Normal_Corneal Thickness, 548 Abnormal_Corneal Thickness, 765 Normal_Elevation Front, 167 Abnormal_Elevation Front, 693 Normal_Elevation Back, and 229 Abnormal_ Elevation Back maps.

*2.2. Transfer Learning Models*

There are numerous common transfer learning models available in computer vision that are typically utilized as a tool for the categorization of medical images; however, in this study, the MobileNetv2 [34], Resnet50 [35], Xception [36], Vision Transformer (ViT) [37], Co-scale conv-attentional image Transformers (CoaT) [38], and Swin transformer (Swin-T) [39] models have been used, which are trained by the original and synthesized images to evaluate the system's effectiveness for diagnosing corneal instances. The models demonstrate the influence of synthesized and imbalanced datasets on the corneal diagnosis task; the data were manipulated, and varied numbers of data were used for training and testing. To be balanced, the data were processed using the resample method (oversampling and downsampling). After training each transfer learning model, the results are compared to the results of other approaches see tables 3 and 4.

The Resnet50 forecasts the delta required to get from one layer to the next and arrive at the final prediction. It addresses the vanishing gradient problem by enabling the gradient to flow through an additional shortcut path. It enables the model to skip over a CNN weight layer if it is not required. This helps to avoid the difficulty of overfitting the training set. ResNet50 is a 50-layer network [36]. The MobileNetv2 is a convolutional architecture built for usage with mobile or low-cost devices that minimizes network cost and size [40]. Segmentation, classification, and object recognition may all be performed with the MobileNetV2 model. In comparison to its predecessor, MobileNetV2 includes two new features occurring linearly between layers, and bottleneck shortcuts are established [41]. Xception is a depthwise separable convolutions-based deep convolutional neural network architecture; Google researchers came up with the idea. Xception has three different flows: entry, middle, and exit. The data initially pass via the entering flow, then eight times through the middle flow, and finally through the exit flow. Batch normalization is applied to all convolution and separable convolution layers [36].

[37] have investigated the possibility of using transformers for straightforward image recognition. Apart from the initial patch extraction step, this architecture does not have any image-specific inductive biases, which sets it apart from previous research leveraging self-attention in computer vision. Instead, [37,42] employs a standard transformer encoder seen in natural language processing to decode an image as a sequence of patches. With pre-training on massive datasets, this straightforward approach scales remarkably well. Therefore, vision transformer competes with or outperforms the state-of-the-art on many picture classification datasets, while only requiring a little initial investment. CoaT, an image classifier based on the transformer, features cross-scale attention and efficient conv-attention operations, and is given in [38]. CoaT models achieve strong classification results on ImageNet, and their utility for subsequent computer vision tasks, such as object detection and instance segmentation, has been established. In [39], a novel vision Transformer called Swin-T is introduced; it generates a hierarchical feature representation and scales computationally linearly with the size of the input image.

For all models, corneal images were fed into the networks to train the models and extract the weights. For 20 epochs, we used a batch size of 32. Moreover, we employed the Adam optimization approach, with a learning rate of 0.001, to iteratively modify network weights. Table 1 displays all of the parameter values utilized by the various classifiers.

**Table 1.** Values of the parameters used in the classifiers.

| Method. | Image size | Parameters |
| --- | --- | --- |
| MobilenetV2 | 224×224 | 3.5M |
| Resnet50 | 224×224 | 25.6 |
| Xception | 299×299 | 22.9 |
| ViT | 128×128 | 36.3 |
| CoaT | 224×224 | 22M |
| Swin-T | 224×224 | 29M |

*2.3. Generating Synthetic Cornea Images*

The diagnostic ratio is negatively affected by a lack of data [43], and this is the fundamental challenge with model training [44]. We synthesized new examples that were learned from existing data examples using a new way of producing synthetic corneal images using generative adversarial networks (GANs) to expand the training data and enhance diagnostic rates. GANs are deep CNN networks that generate new data from previously trained data such as images [45]. For synthesizing labeled images of the cornea, we employed conditional GANs [46]. The structure of the CGAN model used in this work (see Figure 2) is two networks that compete against one another to achieve a common goal, which is to learn the distribution of data $p_{data}$ from samples (images in our work). Whereas in the first network, called the generator G network, an image G(x) is generated, usually from noise shaped by the uniform distribution $P_z$, which is close to the target image, as it produces an image representing the class you want to generate, in addition to noise, to function as an assistant factor that aids the model in synthesizing images that are close to reality. On the other hand, the second network, dubbed Discriminator $D$, tries to discern between real and fake images entered into the network; in other words, the input is $x$, whereas the output is D(x). It compares the image created by the rival network to the actual image. The loss function, shown in equation (1), is optimized to train adversarial networks [47].

$$min_G max_{D=} E_{x \sim Pdata} log D(x) + E_{z \sim Pz}[\log(1 - D(G(z)))] \quad (1)$$

where the $D$ is trained to maximize $D(x)$ for images derived from real data and minimize the D(x) that is derived from not real data. On the other hand, the Generator seeks to trick the Discriminator by generating an image $G(z)$, which calls for maximizing the value of $D(G(z))$. These two networks are still in competition during the training phase, with the Generator attempting to improve its performance to deceive the Discriminator, while the latter distinguishes between the real and fake images.

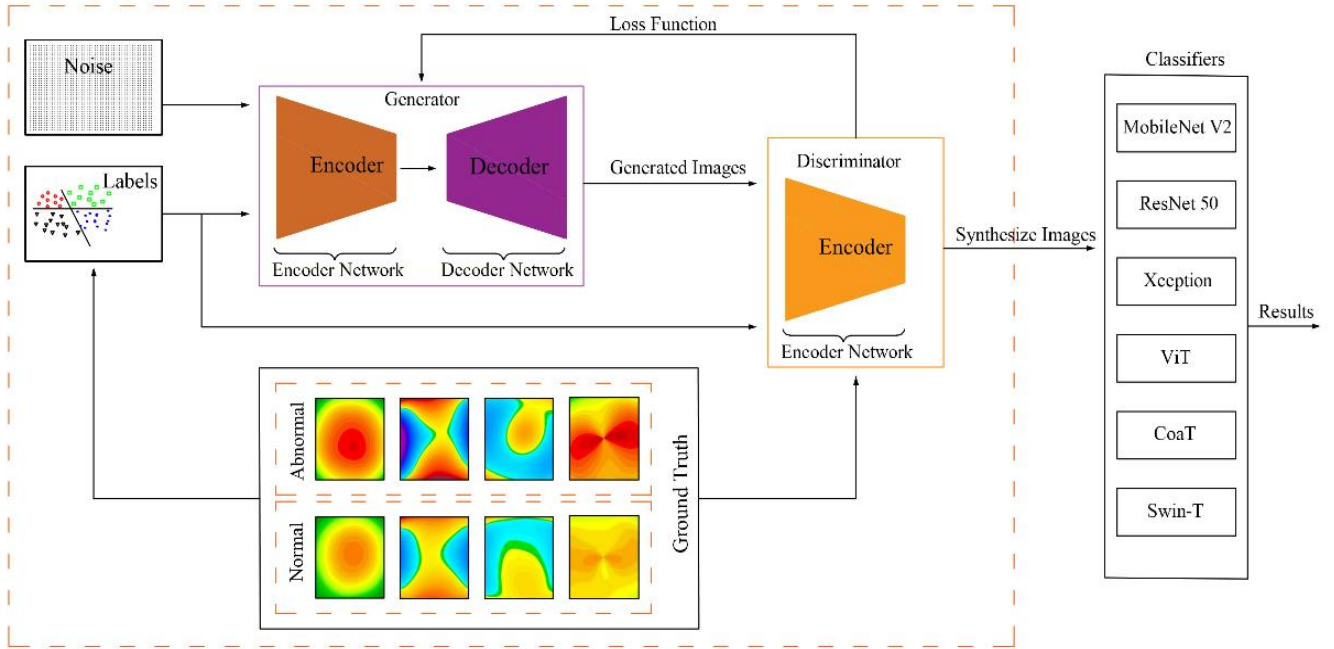

**Figure 2.** *Structure of the proposed system.*

The generator accepts a vector of random numbers with a size of 100 created by uniform distribution, and this vector reshapes into 4×4×1024. The architecture involved four deconvolution layers to up-sample the image using a 5×5 filter size. Finally, the output is the image with a size of 64×64×3. Except for the last layer, batch normalization and ReLU activation functions are used. The Discriminator-issued class label, in addition to the real or fake decision, derives from a corneal image with size 64×64×3 using a filter with size 5x5 with four convolutional layers. To reduce the spatial dimensionality, stride convolution is used in each layer. Batch normalization and ReLU were also applied in each layer (except the fully connected layer).

The training of CGAN was conducted separately to generate every corneal image category, as well as conducted iteratively for the Discriminator and Generator. The noise sample $Z^1 \ldots Z^n$ derives from a uniform distribution in the range [–11], $n = 100$. The slope of the leak of ReLU was equal to 0.2. The zero-centered center normal distribution was employed to initialize the weights with a standard deviation of 0.02. Moreover, for 20 epochs, we used the Adam optimizer, and the learning rate was equal to 0.0001. Figure 2 illustrates the structure of the proposed system.

## 3. Results

The goal of this research, in which all of the steps have been outlined in detail in Table 2, is to find out to what extent generated data affect the diagnosis of corneal diseases, and how well classifiers can classify them. Therefore, the CGAN model has been trained to deal with data disparities; in other words, each corneal disease's image generated is separated with high-quality topographical images by using fine-tuning parameters to disband the scarcity of the cornea dataset. For clinical decision transfer, learning methods have been exploited, where the augmented dataset is used in training the networks.

**Table 2.** Proposed Method:

| | **Inputs: D:** Dataset, **img:** a cornea's image which is selected from the **D**; |
|---|---|
| 1 | *GI = Build a model **M** which generate images from **noise** and targeting **D*** |
| 2 | *For I = 1: CNN classifieres // (MobilenetV2, Resnet50, Xception, ViT, CoaT, and Swin-T)* |
| 3 | ***[accuracy, precision, recall, f1-score]** = Calculate metrics [Accuracy, Precision, Recall, F1-score] from **GI*** |
| 4 | *End for* |
| 5 | *[**SSIM, MSE, PSNR, FID**] = Calculate [**SSIM, MSE, PSNR, FID**] between an image from **GI** and **D*** |
| 6 | *End* |

The results of diagnosing corneal diseases are reported using different types of transfer learning models, such as MobileNetv2, Resnet50, and Xception.

To detect the importance of data generation, as well as its effect on classification tasks, we used the original dataset to train and test each classifier with and without corneal-generated images.

On the other hand, to assess the strength of the synthesis model and its ability to synthesize convergent data in a particular category and divergent from other categories, each classifier was trained on the synthesized data without using the original data. We employed eight-fold cross-validation with case separation at the patient level in all of our experiments and evaluations. The used examples contained the corneal cases (normal or abnormal for each corneal map).

For each batch of data images, we trained the network and assessed the outcomes individually. The CGAN architecture is used to train each corneal-case class separately, utilizing the same eight-fold cross-validation method and data split. Following training, the generator is capable of creating realistic corneal case images separately using a vector of noise formed by uniform distributions (see Figure 3). Accordingly, the model synthesized eight different cases of corneal images: normal and abnormal cases for sagittal, corneal thickness, elevation front, and elevation back images.

We employed two main kinds of metrics in our research. First, we used observational error metrics such as accuracy, precision, recall, and F1-score metrics to evaluate classification accuracy (equations 2, 3, 4, and 5, respectively). Second, we used equations 6 and 7 to evaluate the synthesized image's quality with the original images via the structural similarity index method (SSIM) [48] and the peak signal-to-noise ratio (PSNR) [49].

$$Accuracy = \frac{TP+TN}{TN+TP+FN+FP} \quad (2)$$
$$Precision = \frac{TP}{TP+FP} \quad (3)$$
$$Recall = \frac{TP}{TP+FN} \quad (4)$$
$$F1\_Score = \frac{2TP}{2TP+FP+FN} \quad (5)$$

where TP = true positives, TN = true negatives, FP = false positives, and FN = false negatives.

Structural similarity (SSIM) [48] is an image quality measurement based on equation (6) between the approximated image $y_e^L$ and the ground truth image $y_t^L$.

$$SSIM(y_t^L, y_e^L) = \frac{1}{M}\sum_{j=1}^{M} \frac{(2\mu_{jt}\mu_{je}+c_1)(2\sigma_{jte}+c_2)}{(\mu^2_{jt}+\mu^2_{je}+c_1)(\sigma^2_{jt}+\sigma^2_{je}+c_2)} \quad (6)$$

In contrast, peak signal-to-noise ratio (PSNR) [49] is an objective assessment based on comparisons using particular numerical criteria [50,51]; a higher PSNR value indicates better image quality. Images generated by equation (7) have significant numerical differences at the low end of the PSNR scale [52,53].

$$PSNR(f, g) = 10\log_{10}(\frac{255^2}{MSE(f,g)})) \quad (7)$$

MATLAB2020b is used for the implementation of corneal diagnosing. All training processes were performed using an NVIDIA GeForce GTX 1660 GPU.

Using the above-mentioned metrics for different classifiers, few results were recorded when no synthesized data were used; this might be due to overfitting over the smaller number of training images. Conversely, using the CGAN model, the results improved as the number of training instances grew (see Table 3).

**Table 3.** Performance comparison for classification of corneal conditions among obstetric models (%).

| Classifier | Data | Accuracy | Precision | Recall | F1-score |
|---|---|---|---|---|---|
| MobilenetV2 | Original | 75.2 | 72.4 | 73.2 | 72.3 |
|  | Synthesized | 88.6 | 86.5 | 89.8 | 87.5 |
| Resnet50 | Original | 77.13 | 74.6 | 74.6 | 74.3 |
|  | Synthesized | 90.5 | 90 | 90.4 | 90.1 |
| Xception | Original | 78.9 | 75.6 | 75.7 | 75.1 |
|  | Synthesized | 90.7 | 90 | 90.6 | 90.2 |
| ViT | Original | 71.2 | 68.2 | 68.1 | 67 |
|  | Synthesized | 88.7 | 90.7 | 84.4 | 86.2 |
| CoaT | Original | 65.6 | 64.9 | 65.2 | 65.1 |
|  | Synthesized | 69.3 | 68.1 | 68.4 | 68.2 |
| Swin-T | Original | 58.4 | 56.3 | 57.5 | 56.9 |
|  | Synthesized | 63.4 | 62.5 | 62.7 | 62.6 |

Since our data images are unbalanced, we suggested revealing how the corneal diagnosis would be affected if a balanced dataset was available. Therefore, we used the traditional data balancing methods, where we conducted data resampling using both approaches to make a balanced dataset out of an imbalanced one. The first approach was undersampling (keeping all samples in the rare class and randomly selecting an equal number of samples in the abundant class); the second approach was oversampling (increasing the size of rare samples using repetition). These two approaches were applied to the data before and after generating images.

Results reported that, generally, when applying data resampling on the original data (before using the CGAN model), the classifiers achieved a moral performance, while the data were balanced. Moreover, training by oversampling synthesized data for all classifiers outperforms training by underdamped synthesized data. On the other hand, applying oversampled data on the generated image (after implementing the CGAN model) will not affect the classifier results since the data are vast enough to train the models correctly. In contrast, undersampling negatively affected the achievement of classifiers due to the data being decreased again (see Table 4).

**Table 4.** Performance comparison for classifying corneal conditions among obstetric models after balancing data (%).

| Classifier | Data | Accuracy | | Precision | | Recall | | F1-score | |
|---|---|---|---|---|---|---|---|---|---|
| | | OVS | UNS | OVS | UNS | OVS | UNS | OVS | UNS |
| MobilenetV2 | Original | 85.5 | 75.4 | 85.7 | 76.2 | 85.5 | 75.4 | 85.3 | 75.4 |
| | Synthesized | 88.5 | 81.1 | 88.8 | 81.6 | 88.4 | 81.1 | 88.4 | 81 |
| Resnet50 | Original | 86.36 | 75.7 | 86.3 | 76 | 86.6 | 75.7 | 86.3 | 75.5 |
| | Synthesized | 90.2 | 82.8 | 90.8 | 83 | 90.8 | 82.8 | 90.8 | 82.7 |
| Xception | Original | 86 | 77.3 | 86.2 | 78 | 86 | 77.3 | 85.9 | 76.7 |
| | Synthesized | 90 | 82.7 | 90.3 | 82.7 | 90 | 82.6 | 90 | 82.6 |
| ViT | Original | 74.5 | 70.9 | 73.2 | 68.4 | 72.8 | 69.5 | 73 | 68.9 |
| | Synthesized | 89.8 | 86.1 | 88.2 | 85.5 | 88.9 | 85.6 | 88.5 | 85.6 |
| CoaT | Original | 69.4 | 63.7 | 69.1 | 62.6 | 68.9 | 62.8 | 69 | 62.7 |
| | Synthesized | 73.8 | 66.8 | 72.6 | 65.7 | 72.9 | 65.9 | 72.7 | 65.8 |
| Swin-T | Original | 60.2 | 56.9 | 59.8 | 56.5 | 58.9 | 56.5 | 59.3 | 56.5 |
| | Synthesized | 65.6 | 61.7 | 64.6 | 60.8 | 64 | 60 | 64.3 | 60.4 |

OVS: oversampling, UNS: undersampling.

This issue of whether the set of images generated was sufficiently distinct to allow classification between the corneal case categories was investigated with the help of an expert ophthalmologist. We provided him with 500 randomly generated images with various categories to classify and diagnose. Table 5 summarizes the findings, and Table 6 shows the average of SSIM and PSNR for a random selection of 100 images.

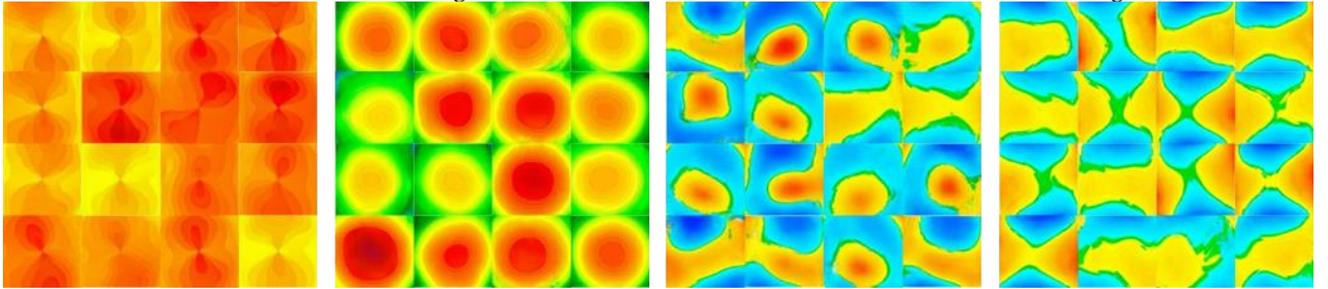

**Figure 3.** Samples of the generated image using the Conditional Generative Adversarial Network (CGAN) model.

**Table 5.** The results from the expert (%).

| | Sagittal Images | CT Images | EF and EB Images |
|---|---|---|---|
| Diagnosis by an Expert | 0.94 | 0.98 | 0.93 |

**Table 6.** Average of Structural Similarity Index (SSIM) and peak signal-to-noise ratio (PSNR) for 100 random images.

| SSIM | PSNR |
|---|---|
| 0.872 | 33.221 |

The SSIM and PSNR have been calculated before and after training the CGAN model on a random sample of 100 images. Table 6 shows that the model can generate synthetic images very close to the original. Therefore, we can consider those images to be legitimate for training CNNs models, and ophthalmologists can use them in clinical research.

The CNN classifiers are repeatedly tested in this work to determine the testing process. The suggested model can be applied in real-time, where testing images only takes a few moments, according to Table 7. While the CoaT model requires the longest ATT, the ATT for ViT beats the other classifiers.

**Table 7.** Convolutional neural network **(**CNN) classifier's average time test (ATT) (sec.).

| MobilenetV2 | Resnet50 | Xception | ViT | CoaT | Swin-T |
|---|---|---|---|---|---|
| 0.0258 | 0.0187 | 0.0152 | 0.0108 | 0.0342 | 0.0203 |

The high quality of the images can be seen in the images synthesized from the test images using the CGAN model, which are displayed in Figure 4. It is also possible to notice the stability of the structures and morphologies of the images.

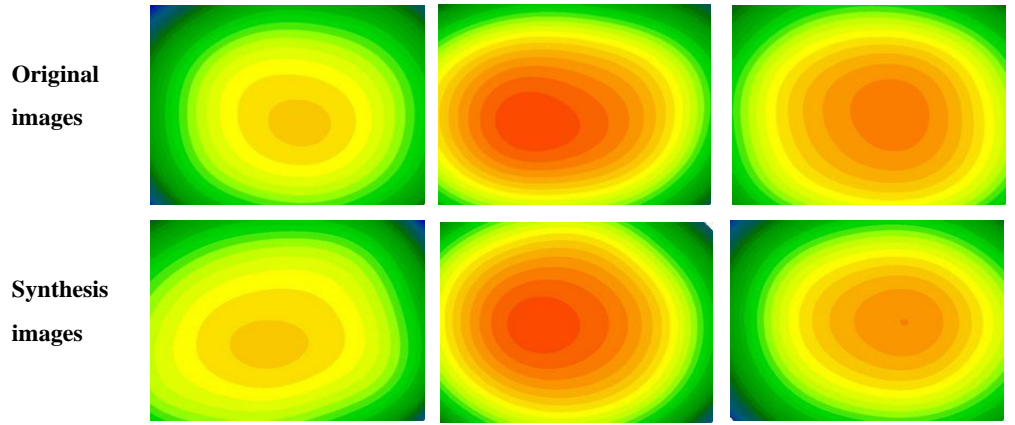

**Figure 4.** Example of original and synthesis images.

## 4. Discussion

The objectives of this work were to apply the CGAN model to generate synthetic medical images for data augmentation to expand limited datasets and improve clinical decision-making for corneal diseases. Thus, we investigated the extent to which synthetic corneal images help another system perform better behind the scenes. The study used a small dataset comprising the sagittal, corneal thickness, elevation front, and elevation back of corneal images. Each class has its distinct characteristics, although there is considerable intra-class variation. Our diagnosis was based on the four maps, each of which was examined to determine whether it was normal or diseased. To identify corneal disorders, a variety of transfer learning architectures were employed. We discovered that by utilizing the CGAN model to synthesize extra realistic images, we could increase the size of the training data groups, thus boosting the clinical decision. The diagnostic outcomes for mobilenetV2, Resnet50, Xception, ViT, CoaT, and Swin-T classifiers improved from 75.2 % to 88.6 %, 77.13% to 90.5%, 78.9% to 90.7 %, 71.2% to 88.7%, 65.6% to 69.3%, and 58.4% to 63.4%, respectively. Results from Table 3 show that the synthetic data samples generated can increase the variability of the input dataset, resulting in more accurate clinical decisions.

The scores demonstrate that the synthesized images have useful visuals and, more crucially, useful characteristics that may be used in computer-aided diagnosis. The other aspect of this research is to test the effect of data balance on diagnostic results, where we used the resampling method to make the dataset balanced. The results showed that training the model before generating a new set of data on a balanced dataset is very important, especially in circumstances where data are scarce. On the contrary, we did not notice a significant impact on the performance of the classifiers when using the data resampling on the generated data because the data was sufficient and suitable for training the models without the need to balance them using data balancing methods. This is clear evidence of the importance of the model proposed in this paper. In a final experiment, we compared the performance of the classifiers-based systems employed in this study for clinical decision-making (Table 4). The highest performance was derived from synthesized data in the Xception classifier, whereas the best performance came from using balance data in Resnet50 when using the oversampling approach, but the ViT model while using the undersampling approach.

This work has several limitations. For example, the training complexity was enhanced by training distinct GANs for each corneal case class. It might be useful to look into GAN designs that produce multi-class samples at the same time. Another type of GAN learning process might increase the quality of the corneal image. It is also possible to do more research to improve the training loss function by adding regularization terms.

Because the human factor is critical in evaluating the proposed model's outputs, an expert opinion was obtained after providing him with a set of generated corneal images containing a randomly selected set of normal and abnormal corneal images. The following was the expert's opinion: "Creating a new template for the corneal topographical of four refractive maps is considered an interesting subject as it enriched the overall expected shapes that could be seen during the daily clinic. These new images which created based on real cases collected previously and diagnosed that the new images are still inside the reality borderlines. Gain good experience with the new shapes and specify the further required steps of a diagnosis other than the topographical maps that could be specified advanced for predicted out-of-skim cases. In such a way, offline training for the new ophthalmologists and improving the skill of diagnosis with the preparation for new unseen cases could be done." In the future, we look to develop our research to exploit other GANs that might benefit from corneal image synthesis for better achievement.

## 5. Conclusion

In conclusion, we proposed a strategy for improving performance in a medical issue with little data by generating synthetic medical images for data augmentation. On a corneal diseases diagnosis task, we discovered that synthetic data augmentation beat traditional data augmentation in accuracy by roughly 13%. Additionally, we investigated the performance of the classifiers in different conditions, and we found that while working with cornea images to diagnose diseases, the Xcepton classifier is more responsive than the rest of the used classifiers. We anticipate that synthetic augmentation can help with a variety of medical issues and that the method we have outlined can lead to more powerful and reliable support systems.

**Author Contributions:** "Conceptualization, S. K. J. and S. A.; methodology, S. K. J., S. A., N. H. G., and J. M.; software, S. K. J. and S. A.; validation, S. A., N. H. G., and J. M.; formal analysis, S. K. J., and S. A.; investigation, N. H. G., and J. M.; resources, S. K. J., S. A., N. H. G., and J. M.; data curation, S. K. J., S. A., N. H. G., and J. M.; writing—original draft preparation, S. K. J., S. A., N. H. G., and J. M.; writing—review and editing, T. A. R. and S. Q. S.; visualization, T. A. R.; supervision, T. A. R.; project administration; funding acquisition, S. Q. S. and P. S. J. All authors have read and agreed to the published version of the manuscript."

**Funding:** Dr.P. S. JosephNg, Faculty of Data Science & Information Technology, INTI International University, Persiaran Perdana BBN, 71800 Nilai, Negeri Sembilan, Malaysia.

**Institutional Review Board Statement:** The manuscript is conducted within the ethical manner advised by the targeted journal.

**Informed Consent Statement:** Not applicable.

**Data Availability Statement:** Data can be shared upon request from the corresponding author.

**Acknowledgments**: None.

**Conflicts of Interest:** The authors declare no conflict of interest to any party.